\documentstyle[aps,prl,preprint,epsfig,amsmath,tighten,graphicx]{revtex}
\begin{document}
%\begin{frontmatter}
\title{Molecular Spintronics: Spin-Dependent Electron Transport in
Molecular Wires}
\author{Eldon G. Emberly$^1$ and George Kirczenow$^2$}
\address{$^1$NEC Research Institute, 4 Independence
Way, Princeton, New Jersey 08540 \linebreak
$^2$Department of Physics, Simon
Fraser University, Burnaby, British Columbia, Canada, V5A 1S6}
\date{\today}
\maketitle
\begin{abstract}
We present a theoretical study of spin-dependent transport through
molecular wires bridging ferromagnetic metal nanocontacts. We extend
to magnetic systems a recently proposed model that provides a {\em
quantitative} explanation of the conductance measurements of Reed {\em
et al.}\cite{Reed1} on Au break-junctions bridged by self-assembled
molecular monolayers (SAMs) of 1,4-benzene-dithiolate (BDT)
molecules. Based on our calculations, we predict that spin-valve
behavior should be observable in nickel break-junctions bridged by
SAM's formed from BDT. We also consider spin transport in systems
consisting of a clean ferromagnetic nickel STM tip and SAMs of
benzene-thiol molecules on gold and nickel substrates. We find that
spin-valve behavior should be possible for the Ni substrate. For the
case where the substrate is gold, we show that it should be possible
to inject a highly spin-polarized current into the substrate.
\end{abstract}

%\begin{multicols}{2}
%\end{frontmatter}
\section{Introduction}
Over the past few years, molecular wires formed by bridging pairs of
metallic contacts with organic molecules, have been shown to display a
remarkable range of electron transport
phenomena\cite{Reed1,Datta,Metzger,Collier,Joachim,Reichert,Reed2,Schon}
Some wires display rectification\cite{Metzger}, others yield negative
differential resistance\cite{Reed2} and/or transistor
action\cite{Schon}, and still others involve molecules that are known
to switch conformations and exhibit switching behavior in their
transport characteristics\cite{Collier}.  As this special issue on
electron transport will attest, understanding the measured
experimental data is a great challenge that requires bringing together
principles from both chemistry and physics. A wide variety of
theoretical approaches have arisen over the past few years for the
modeling of molecular wire systems. These range from self-consistent
calculations\cite{Lang,DiVentra,Guo,Ratner,Emberly3,Emberly4,Damle,Kosov,Gutierrez}
to semi-empirical
methods\cite{Joachim2,Yaliraki,Emberly1,Hall,Emberly5}.  Molecular
wires have great promise from a pure science perspective and also
potentially for applications. Hence both experimental and theoretical
research in this field continues to attract increasing interest.

Another area of research that has also seen great growth over the past
few years is spintronics, a branch of electronics that employs the
electron's spin degree of freedom as well as its charge to store,
process and transmit
information\cite{Prinz,Kikkawa,Datta-Das}. Important spintronic
phenomena include giant magnetoresistance, spin valve behavior and
injection of spin polarized currents from ferromagnetic materials to
paramagnetic
materials\cite{Mes,Julliere,J&S,GMR,Alvarado,Fiederling,Ohno,Schmidt,Rashba,Kirczenow,LaBella,Zhu}.
The prototypical spin valve system consists of two ferromagnetic leads
separated by a paramagnetic coupling layer. Transport is measured
between the two ferromagnetic layers under an applied voltage. The
system functions as a spin valve when it exhibits a difference in its
electrical conductance between configurations in which the
magnetizations of the ferromagnetic layers are aligned parallel and
anti-parallel.  Different materials can display very different
spintronic properties. For example, injection of spin currents from
metallic ferromagnets into non-magnetic metals is relatively straight
forward and was first demonstrated experimentally more than a decade
ago\cite{J&S}. Since then it is has found important practical
applications in information storage technology\cite{Prinz}. However
attempts at demonstrating efficient spin injection from ferromagnetic
metals into semiconductors through solid state interfaces have
encountered fundamental obstacles\cite{Schmidt} and work is underway
to determine whether these may be overcome through the use of
potential barriers and/or interfaces that obey certain selection
rules\cite{Schmidt,Rashba,Kirczenow,LaBella,Zhu}.

Much less is known at the present time about spin transport in
molecules. There have been a few studies of spintronic systems
involving carbon nanotubes\cite{Tsukagoshi,Mehrez}. There has also
beeen theoretical work looking at transport through paramagnetic ions
bridging two electrodes in the presence of a magnetic
field\cite{Petrov}. However to our knowledge whether
functioning self-assembled nanoscale spintronic devices can be made
out of organic molecules and magnetic nanocontacts has not as yet been
explored experimentally or theoretically. In this paper we show that
spin-dependent transport phenomena should be readily observable in
such molecular wire systems using currently available experimental
techniques. We consider two classes of systems: break-junctions made
from ferromagnetic nickel bridged by self-assembled monolayers of
organic molecules, and STM setups consisting of a clean single-crystal
nickel tip and a SAM on a gold or nickel (100) substrate. Schematics
of these systems are shown in Fig.~\ref{fig1}. Our calculations
suggest that spin-valve effects and spin injection should be possible
in these molecular wire based systems.

As a reference point for the present calculations, we briefly revisit
a recently proposed model that has been instrumental in achieving a
quantitative understanding\cite{Emberly5} of current flow in
self-assembled molecular wires bridging a non-magnetic break-junction.
In this model it is assumed that current flows not through a single
molecule connecting the two tips of the metal break-junction as in
previous theories, but through overlapping molecules, with each
molecule chemically bonded to only one metal tip. This model together
with a semi-empirical method for the solution of the transport problem
is able to quantitatively reproduce\cite{Emberly5} the measured
conductance data of Reed {\em et al.}\cite{Reed1} for gold break
junctions bridged by benzene-dithiol molecules. Because it is not
compute-intensive, a semi-empirical theoretical approach allows us to
calculate the transport properties of an ensemble of possible
orientations of the overlapping molecules and to include the effects
of positional disorder. Thus we are able to calculate transport
properties averaged over configurations. We feel that the success of
this work has demonstrated that semi-empirical methods are useful as a
tool for evaluating and estimating effects that are not easily
accessible to ab-initio techniques.

This paper is organized as follows: In Sec.~2, we give a brief
summary of a method for calculating the transport characteristics of a
molecular wire. The current is evaluated via Landauer theory and the
scattering problem is solved using a semi-empirical tight-binding form
of Schroedinger's equation. Sec.~3 contains a summary of the model
that we have proposed for understanding electrical conduction in
non-magnetic metal break-junctions bridged by SAM's\cite{Emberly5},
and also representative results of calculations of the differential
conductance of non-magnetic systems that are in quantitative agreement
with experiment\cite{Reed1}. These results provide a useful benchmark
for comparison with our calculations for magnetic systems that
follow. Our spin-dependent transport results are presented in Sec.~4
and 5. In in Sec.~4 we consider molecular wires bridging magnetic
break junctions. We predict that spin-valve behavior should be
experimentally detectable in break-junction measurements on SAM's,
where the break junction is made from ferromagnetic nickel. Sec.~5
details our results for systems that include magnetic STM tips. We
show that spin-injection and spin-valve effects should be observable
in the systems that we consider. Finally, in Sec.~6 we present some
conclusions that may be drawn from this work.

\section{A method for evaluating the electrical current in molecular wires}
In this section, we briefly review a method for calculating the
conductance properties of a molecular wire system. We consider a
molecular wire to consist of a pair of electrodes which act as a
source and drain for electrons, that are connected by a molecule or
molecules. A bias is applied to the electrodes and an electric current
flows through the system. A standard method for evaluating the current
in such a system is Landauer theory\cite{Landauer}. It relates the
electric current flowing through the molecular conductor to the
probability for a single electron to scatter from one electrode to the
other.  For a two-terminal system, the current as a function of the
applied bias $V$ can be calculated from the following equation
\begin{equation}
I(V) = \frac{e^2}{h} \int dE\: T(E,V) [F(E,\mu_S) - F(E,\mu_D)]
\label{eq:landauer}
\end{equation}
where $F(E,\mu)$ is the equilibrium Fermi distribution and $\mu_{S,D}
= E_F \pm eV/2$ are the electro-chemical potentials of the electrodes
in terms of the
common Fermi energy $E_F$ of the source (S) and drain (D) electrode
respectively. The transmission probability $T(E,V)$ is the sum of the
transmission probabilities for both spin up and spin down electrons to
scatter through the system at energy $E$. For a thorough review of
Landauer theory see Ref.~\cite{Dattabook}.

A variety of methods exist for solving the scattering problem to
determine the transmission probability. These range from
self-consistent calculations based on density-functional
theory\cite{Lang,DiVentra,Guo,Ratner,Emberly3,Emberly4,Kosov}, to simpler
semi-empirical tight binding
approaches\cite{Joachim2,Yaliraki,Emberly1,Hall,Emberly5}.
The density-functional methods have the advantage of
self-consistently calculating the effects of the
electro-static potential that arises in the vicinity of the
molecule due to the applied bias $V$ and any charge transfer
that may arise between the leads and the molecule. They have
the disadvantage of being numerically intensive which
restricts the size of the problems to which they can be
applied at present. They constitute the present state of the
art, but never-the-less they rely on approximations whose
reliability for transport calculations is unknown when the
system departs significantly from equilibrium, a case that
is particularly important for molecular wire systems.
Semi-empirical approaches do not include the effects of the
electro-static field self-consistently.  Many-body effects
are included implicitly to a limited extent in the
parameters, and the electro-static field may be added as an
extra potential which can be used to approximate the true
field. Because of the speed with which calculations using
semi-empirical methods can be carried out, they have the
advantage of being able to study much larger systems,
including also the various changeable degrees of freedom
that influence transport in these systems.  They have been
shown to produce quantitatively similar results to density
functional calculations in the linear voltage regime for
systems where the effects of the self-consistent potential
are less significant\cite{Emberly1,DattaBDT,Emberly2}. Thus
semi-empirical approaches should be considered as a valuable
tool in that they are able to provide an efficient method by
which to evaluate the importance of various factors which
may contribute to the transport properties of the molecular
wire of interest. With the important factors determined,
more sophisticated approaches can then be used to gain more
quantitative understanding. What follows is a summary of the
semi-empirical tight-binding approach that we use in this
work to solve the scattering problem.

We calculate the transmission probability by explicitly calculating the
single electron scattering states. Our starting point is
Schroedinger's equation
\begin{equation}
H |\Psi\rangle = E |\Psi\rangle \label{eq:schroed}
\end{equation}
where $|\Psi\rangle$ is the single-particle scattering state at energy
$E$ and $H$ is the Hamiltonian for the entire system. For a two
terminal molecular wire, we consider $H$ to have the form
\begin{equation}
H = H^0_L + H^0_M + H^0_R + V_L + V_R + V_{ext} + V_{SCF}
\end{equation}
where $H^0_{L,M,R}$ are the Hamiltonians for the isolated left lead
(L), molecule (M) and right lead (R). The molecule is coupled to the
leads via $V_{L,R}$ for the left and right leads
respectively. $V_{ext}$ is the external potential due to the applied
bias, and $V_{SCF}$ represents the self-consistent electro-static
potential which depends upon the charge distribution in the
system. (In what follows, we no longer consider this
term, however a tight-binding method including this term is given in
Ref.~\cite{Emberly3,Emberly4}). The scattering state $|\Psi\rangle$ can be
written as $|\Psi\rangle = |\Psi_L\rangle + |\Psi_M\rangle +
|\Psi_R\rangle$, with the following conditions
\begin{eqnarray}
|\Psi_L\rangle &=& |\Phi^\alpha_{+,L}\rangle + \sum_{\alpha'}
     r_{\alpha',\alpha} |\Phi^{\alpha'}_{-,L}\rangle \\
|\Psi_M\rangle &=& \sum_i c_i |\phi_i\rangle \\
|\Psi_R\rangle &=& \sum_{\alpha'} t_{\alpha',\alpha}
|\Phi^{\alpha'}_{+,R}\rangle
\end{eqnarray}
where $|\Phi^{\alpha'}_{\pm,L/R}\rangle$ is the ${\alpha'}^{th}$ mode
at energy $E$ in either the left or right leads, with the $\pm$
denoting whether the state is a rightward or leftward propagating or
evanescent mode. The $\alpha^{th}$ mode is always a rightward
propagating mode. The $|\phi_i\rangle$ are the bound eigenstates of
the isolated molecule. The $r_{\alpha',\alpha}$ and
$t_{\alpha',\alpha}$ are the reflection and transmission coefficients
respectively. The total transmission probability for an electron to go
from the left lead to the right lead is given by
\begin{equation}
T(E,V) = \sum_{\alpha,\alpha'}
\frac{v_{\alpha'}}{v_\alpha}|t_{\alpha',\alpha}|^2
\end{equation}
where the sum over $\alpha$ is over rightward propagating modes in the
left lead and the sum over $\alpha'$ is over rightward propagating
modes in the right lead; all modes at energy $E$. If the system is
magnetic, we assume the effects of spin orbit coupling on transport
through the molecular wire to be negligible\cite{so} and solve for
$T(E,V)$ for the spin up and spin down electrons separately.

We solve the above problem by representing the scattering states using
a linear combination of atomic orbitals (LCAO) (also known as the
tight-binding approximation)\cite{Huckel}. This yields a matrix
form of Eqn. {\ref{eq:schroed}
\begin{equation}
\sum_j H_{ij} \Psi_j = E \sum_j S_{ij} \Psi_j
\label{matrix}
\end{equation}
where $H_{ij}$ is the Hamiltonian matrix in the chosen atomic orbital
basis.  If the basis of atomic orbitals is non-orthogonal then the
non-diagonal elements of the overlap matrix $S_{ij}$ must be included;
an exact and easily implemented way to take them into account in
molecular transport calculations is described in
Ref.\cite{Hilbert1,Hilbert2} Eq. \ref{matrix} is a linear system which
can be solved in a straightforward manner for the unknowns
$r_{\alpha',\alpha}$, $c_i$, and $t_{\alpha',\alpha}$. The various
propagating and evanescent modes in the leads are determined using
standard transfer matrix methods.

We will use this method to explore spin transport in systems that include
ferromagnetic leads below. However, we first discuss our model in
the simpler context of a non-magnetic system for which
it provides a quantitative explanation of available
experimental data. This discussion will serve as a theoretical
and experimental benchmark that will be helpful in understanding
the predictions for magnetic systems that
will follow.

\section{Transport in non-magnetic metal break-junctions bridged by
self-assembled monolayers of organic molecules} In this section, we
present a summary of a model\cite{Emberly5} which can quantitatively
account for the data measured by Reed {\em et al.}\cite{Reed1} on a
gold break-junction bridged by self-assembled monolayers (SAMs) of 1,4
benzene-dithiolate (BDT).  Measurements on this system at room
temperature displayed an apparently insulating region for bias values
between approximately -0.7 and 0.7 volts, nearly symmetric
differential conductance in the range from -4 to +4 volts and some
asymmetry at higher bias.  The measured conductance was much lower
than the quantum $2e^2/h = (1/12.9$k ohm) that sets the conductance
scale for single-channel metallic nanowires and also much lower than
the calculated conductances of molecular wires in which a single BDT
molecule bridges the gap between the two
contacts\cite{DattaBDT,Emberly1,DiVentra,Damle,Hall}; the first experimental
conductance plateau occurred at a resistance of 22M ohm.

In the present model, instead of a single molecule chemically bonding
to both leads, we consider the situation where each molecule bonds to
only one metal tip and the current flows through overlapping molecules
from one lead to the other; Fig.~\ref{fig1}a is a schematic
of the model we consider.  It has been previously shown
theoretically, that having neighboring molecules through
which current can flow can have important consequences
\cite{Yaliraki2,Langpara}.  The present model accounts for the
observed low conductance\cite{Reed1} as a result of the weak
electrical coupling between the two overlapping molecules even though
there is strong coupling between each molecule and the lead to which
it is attached. It can also account for the observed reproducibility
of the measurements\cite{Reed1} after repeated separation of the two
tips: no chemical bonds are formed between the overlapping molecules
or between the two leads, hence the SAM's can be pulled apart and
brought back together reversibly.  Regarding the observed symmetric
behavior of the differential conductance \cite{Reed1}, we argue that
this symmetry arises from the system having the ability to sample all
possible energetically favorable configurations of the overlapping
molecules. Hence even though a given configuration of the two
molecules may be asymmetric, because of thermal fluctuations the
mirror configuration will also be sampled. Hence the average over the
transport properties of all configurations yields a symmetric result,
the small residual asymmetry of the current voltage characteristic in
the experiment\cite{Reed1} being due to differences between the atomic
structures of the two tips.

Since both tips are densely coated with molecules we assume that
molecules bonded to one tip can not also bond to the other. Because of
this, the ends of the molecules that do not bond chemically to a gold
surface remain as thiols (SH).  For simplicity we consider the case
where the current is carried predominantly through only a single pair
of overlapping molecules. We argue that on average the faces of the
two benzene rings should be oriented perpendicularly to each other
because of electrostatic effects. The rings of carbons in similar
molecules have been shown to have negative charge on their sides,
whereas the hydrogens all have positive charge\cite{Jung}.  Thus a
favorable configuration of two rings which minimizes the electrostatic
energy should be one in which they are oriented perpendicularly to
each other; this has been found to give rise to herring-bone patterns
in molecular dynamics simulations of SAM's of benzene-thiol and
similar molecules on gold\cite{Jung}.  Fig.~\ref{fig1}a shows a
possible configuration, where each molecule is bonded to the end of
one tip and the molecules orient themselves to avoid steric clashes
and minimize the electrostatic forces between them.

We simulate the above system by representing the break-junction by
gold clusters to which we attach two BDT molecules.  The gold clusters
are in the [100] direction and consist of 5x5, 4x4, 3x3, and 2x2
layers of atoms, forming 54 atom clusters which form the tips of the
metal break-junction. We only include the gold 6s orbital as we have
found that including all of the orbitals (6s,6p,5d) yields
quantitatively similar results for the bonding geometries considered
here. The BDT molecule consists of a benzene ring with the 1 and 4
hydrogens replaced with sulfur atoms. The sulfur atom that is not
bonded to a lead remains a thiol.  The sulfur atoms that bond to the
gold are positioned over a four-fold hollow site 2 Angstroms from the
ends of the gold clusters\cite{Sellers}. We use the standard extended
H\"{u}ckel parameters\cite{Alvarez} to represent the valence orbitals
of each atom in the system, and evaluate the necessary site energies,
hopping energies and overlaps that make up the Hamiltonian and overlap
matrices that go into the solution of Schr\"{o}dinger's equation,
Eq. \ref{eq:schroed}. We attach to this molecular junction,
semi-infinite 6x6 simple cubic multi-mode leads to act as source and
drain. Each atom in the lead is simulated using just a single
gold-like s orbital. We include disorder by randomly displacing each
atom in the gold clusters by a distance consistent with the
Debye-waller factor for gold. For the BDT molecule, we consider random
displacements of each atom in the plane of the ring. We do this to
approximate the vibrational disorder that will be present in the above
systems at room temperature. For the overlapping ring system, we
consider one molecule to assume a random tilt angle (between 20 and 40
degrees) with respect to the normal from the gold surface and then
orient the second BDT molecule so that its ring is oriented
perpendicular to the first, allowing for 20 degree fluctuations. We
impose steric constraints by demanding that the atoms of the two BDT
molecules be separated by distances $\ge$ 3 Angstroms. The
transmission probability shown below is the average of transmission
probabilities from 50 different atomic configurations.

In Fig.~\ref{fig2}a, the average transmission probability is shown as
a function of incident electron energy. There are several broad
resonances. These resonances are due to highly hybridized
molecule/gold states, with the resonances below -10 eV being due to
the molecular HOMO states and those above -9 eV arising from the LUMO
states. The resonances at -8.2 eV are due to the LUMO $\pi*$ state
which is localized on the benzene ring.  The multiple resonances at
the LUMO are due not only to vibrational disorder but also to the fact
that the wave function overlap between the two molecules results in a
splitting of the almost degenerate LUMO levels. In the limit of
averaging over an infinite number of configurations these peaks would
smear into one broad peak \cite{Transmission}.

The solid curve in Fig.~\ref{fig2}b shows the calculated differential
conductance at room temperature for the case of overlapping molecules
using a Fermi energy for the gold leads of -10 eV. The first rise in
conductance is due to resonant transmission between the HOMO of the
molecules. The 2nd rise can be attributed to transmission between the
LUMO state and also the lower HOMO state. The qualitative and
quantitative agreement with the experimental data of Reed {\em et
al.}\cite{Reed1} (the dashed curve in Fig.~\ref{fig2}b) in the voltage
range below 2 volts is quite striking: The features in the calculated
differential conductance curve appear at nearly the same values of the
bias as in the experimental data, and also the magnitude of the
calculated conductance is close to that which was observed in the
experiment. Above 2 volts the qualitative agreement is less striking,
yet the overall magnitude is in quantitative agreement with the
measured values. In the non-linear regime the effects of
electro-static potential are undoubtedly important and contribute to
the shifting of the molecular orbitals. This would lead to shifting of
the resonances shown in Fig.~\ref{fig2}a. We have attempted to address
qualitatively what this shifting may look like by introducing
potential drops at the contacts between the leads and molecules as
well as in between the two molecules. We assumed that 75\% of the
voltage drops between the two molecules while the other 25\% drops at
the metal contacts (this is a simple assumption which may or may not
reflect the true potential profile of the system). We found that the
resonances do shift somewhat but that the overall magnitude of the
resonances does not change appreciably. To obtain more accurate
results at the higher biases, more sophisticated approaches which are
based on density-functional theory should be used. Other improvements
might be to use monte-carlo techniques to achieve a proper Boltzmann
sampling of the possible configurations of the two molecules.

\section{Spin valve behavior in a ferromagnetic break-junction bridged
by self-assembled monolayers of an organic molecule}
In this section we explore the transport properties of a
break-junction bridged by SAMs of organic molecules where the
break-junction is made of ferromagnetic nickel. Ferromagnets bridged
by conducting layers are of interest from a device standpoint in part
because of the possibility of spin-valve behavior. A spin-valve is
characterized by a change in the conductance of the system when the
magnetizations of the two ferromagnetic layers switch between parallel
and anti-parallel configurations.

To gain some intuitive insight into this behavior, it helps to
consider the simplified band structure of ferromagnetic nickel shown
schematically in Fig.~\ref{fig3}. Nickel is a ferromagnet because it
has unequal spin up and spin down populations. The d-band for spin up
electrons is completely filled, while for spin down electrons it is
partially filled. This gives nickel a net spin which implies a net
magnetization even in zero applied magnetic field. Now consider an
experiment that measures the conductance between two nickel layers. In
the situation where the two layers have their magnetizations aligned
parallel, spin down electrons at the Fermi energy in the d-band of one
layer will be transported into both the d-band and the s-band of spin
down electrons of the other layer (ignoring spin-flip scattering and
assuming that there is a conducting channel between the two
layers). Then consider the situation where the magnetizations are
aligned anti-parallel. Now the spin down electrons at the Fermi level
in the d-band of one layer will be transported into the s-band of the
spin up electrons of the other layer, since there is no d-band for
spin up electrons in the drain at the Fermi energy. This simple
argument implies that the band mismatch in the anti-parallel
configuration will lead to a different conductance. Hence one might
expect a difference in conductance between the two configurations.
Non-magnetic metals, thin insulating layers and vacuum tunnel barriers
are among the systems that have been used successfully as conduits
between ferromagnetic metal layers in such spin valve
experiments. Here we consider the conduit to consist of self-assembled
monolayers of organic molecules.

The structure of the system we consider is similar to that of the
system studied in the preceding section but here the nanoscale
contacts are of (100) nickel.  Again they are in the form of clusters
built from 5x5, 4x4, 3x3 and 2x2 layers of atoms. We also make the
assumption that each cluster forms a single magnetic domain. We again
consider BDT to be the molecule forming the SAM on each tip. We assume
that each sulfur atom that binds to a tip is situated 1.7 Angstroms
above a four-fold hollow site of nickel. The tight-binding parameters
that we use for nickel's 4s,4p and 3d orbitals are taken from
Ref.~\cite{papaconstantopoulos_book_1986} for bulk ferromagnetic
nickel. There are two sets of parameters, one for spin up electrons
and one for spin down electrons.  The parameters for the atoms making
up the molecules are taken from the standard extended H\"{u}ckel set
\cite{Alvarez}. We evaluate the Hamiltonian matrix elements coupling
the nickel tips to the molecules using extended H\"{u}ckel where now
the Ni parameters are the standard ones used in extended H\"{u}ckel
calculations \cite{Alvarez}. Thus we make the assumption that spin
down and spin up electrons interact in the same manner with the
molecule. The energy scale used in
Ref.~\cite{papaconstantopoulos_book_1986} is offset relative to that
used in extended H\"{u}ckel theory. Thus to achieve consistency
between the two schemes we added a constant $\epsilon= -17.2$ eV to
all of the site energies taken from
Ref.~\cite{papaconstantopoulos_book_1986} and also adjusted the
off-diagonal Hamiltonian matrix elements accordingly; for further
details of this procedure see the Appendix. (The shift was calculated
by aligning the Fermi energies: $E_F = 8.7$~eV from
Papaconstantopoulos and $E_F = -8.5$~eV for bulk Ni as calculated from
the parameters used by extended H\"{u}ckel). Again, 6x6 simple cubic
semi-infinite ideal leads were coupled to the last layer of each
cluster, with a Ni s orbital on each site. The coupling between the
ideal leads and the tips were evaluated using extended H\"{u}ckel.

We performed transport calculations for the cases where the
magnetizations of the single domain tips are aligned parallel and
anti-parallel. The transmission probabilities for these two cases are
shown in Fig.~\ref{fig4}. Because the dimensionality of the
Hamiltonian matrix has grown nine fold over the case where the tips
were made of gold, we did not perform averaging over the orientations
of the two molecules -- what is shown is $T(E)$ for only one
particular configuration. (Our purpose here is not so much to make
quantitative predictions, but simply to use the semi-empirical
approach to gain insight into what might be observed since, as was
shown in the preceding section, calculations such as this can make
reasonable assessments about the magnitude of transmission and also
the hybridization that occurs between the molecules and the
metal). Fig.~\ref{fig4}a shows the total transmission for the case
where the tips' magnetizations are aligned parallel. Two calculations
were performed to generate this curve. The first involved calculating
$T_{down}(E)$ for the system where spin down parameters are used on
both tips, and then secondly calculating $T_{up}(E)$ for the case
where spin up parameters are used on both tips. The total transmission
is then $T(E) = T_{up}(E) + T_{down}(E)$. Above $-8.5$ eV, the
dominant channel in Ni is s-type, and hence $T(E)$ displays similar
features to those seen for the gold break-junction system: There is a
resonance at $-8.2$ eV from the LUMO and a broader peak at higher
energy from the next LUMO. Below $-8.5$ eV we see the effects of the
d-states of Ni. The HOMO and LUMO hybridize with the d-states of Ni
near the Fermi energy to form transporting states throughout the
nominal HOMO-LUMO gap of the molecule. Fig.~\ref{fig4}b shows $T(E)$
for the case where the tips are aligned anti-parallel. This curve was
calculated by using spin down parameters on one tip and spin up
parameters on the other. This yields $T_{down\rightarrow up}(E)$ and
due to detailed balance the total transmission is $T(E) = 2
T_{down\rightarrow up}(E)$.  Because of the band mismatch, there are
some differences between the parallel and anti-parallel
configurations. The first resonance below $-8.5$~eV is clearly lower
in amplitude for the case where the spins are aligned anti-parallel
compared to the parallel case. This should manifest itself as a
difference in conductance between these two configurations.

In Fig.~\ref{fig5} we show the calculated differential conductance for
the break-junction system at room temperature using a Fermi energy of
$-8.5$~eV. The solid line corresponds to the case where the two tips
have their magnetizations aligned parallel. One might imagine
separating the tips and applying a magnetic field so as to flip the
spin on one of the tips. Alternatively the break junction may be
fabricated so that the Ni film on one side of the break junction is
thinner than on the other; the magnetization of the thinner film will
reverse at a lower applied magnetic field than that of the thicker
film. The dashed line shows the differential conductance for the
situation where the tips are aligned anti-parallel. The difference in
conductance is not dramatic, although it should be within the realm of
experimental detection. Thus this simple model suggests that spin
valve behavior should observable in a break-junction measurement.

\section{Spin dependent transport in STM measurements on organic molecules}

In this section we consider some possible experiments which utilize
STM apparatus with an STM tip made from single crystal nickel with a
single magnetic domain.  Specifically we consider two systems: one
with a SAM grown on a substrate of gold and secondly one where the SAM
is grown on nickel. We consider the SAM to be made from benzene-thiol
(BT), where the thiol binds to the substrate. The end of this molecule
that does not carry the thiol group should not bond chemically to the
Ni STM tip, and we will assume that the tip is clean, i.e., free from
adsorbed molecules. For the system with the gold substrate we assume
the possibility of detecting any spin-polarized current that may be
injected into it from the STM tip. For the case where the substrate is
nickel, we are again interested in the possibility of spin-valve
behavior.

Fig.~\ref{fig1}b is a schematic of the set up that we consider. The
substrate is taken to consist of (100) metal, and we model it using a
rounded tip consisting of 5x5, 4x4, 3x3, and 2x2 layers of atoms. (As
an aside, we found that using a larger approximately cubic cluster to
model the substrate resulted in dimensional resonances of the cluster
itself being interspersed with the transmission features that concern
us here that are due to the molecular wire and its coupling to the
metal contacts. Using the rounded tip removes these dimensional
resonances from the energy range of interest, and the resonances
presented below are robust to changes of the ideal leads, the coupling
of leads to the clusters, and disorder). The tip is also taken to be
in the [100] direction, and consists of 5x5, 4x4, 3x3, 2x2 layers of
atoms which are capped with a single atom. The molecule is assumed to
be perpendicular to the surface with the sulfur atom situated over a
four-fold hollow site on the substrate.  For the case where the
substrate is gold we assume a binding distance of 2 Angstroms, and
when it is nickel, a distance of 1.7 Angstroms. We assume that the STM
tip is situated over the principal axis of the molecule at a distance
of 2 Angstroms from the end hydrogen atom. We use only the 6s orbital
for the gold atoms in the substrate. For the tip, since there is now a
single atom forming the end of the tip, we use the full valence
orbital set of gold.  This is done because of symmetry considerations:
The s orbital at the end of the tip is orthogonal to the $\pi$ states
of the molecule, whereas the p and d states are not. For systems where
the molecule is coupled to only a single atom we have found that
including the p and d orbitals leads to different transmission
characteristics from the case where only the s orbital was used; we
again take the parameters from extended H\"{u}ckel.

In the case where both the substrate and STM tip are Ni we again use
Papaconstantopoulos' \cite{papaconstantopoulos_book_1986} parameters
for both the spin up and spin down bands adjusted for consistency with
the molecular extended H\"{u}ckel parameters as was discussed in
Sec.~4.  In the case where the STM tip is Ni and the substrate is
gold it is necessary to address the further complication that although
the tip and substrate are of different metals, the Fermi levels (or,
in more precise terms, the electro-chemical potentials) of the tip and
substrate must be equal at zero bias if the two metals are in
electrical contact with each other through the molecular wire. (If
their Fermi levels were not aligned at zero bias then an electric
current would flow until an equilibrium state were reached in which
the electrostatic potential due to the charge transferred between the
two metals would align their Fermi levels and the current would cease
to flow.) This Fermi level alignment is taken into account in our
calculation by applying an equal shift to the site energies and Fermi
level of Ni given in Ref.~\cite{papaconstantopoulos_book_1986}, the
shift (-18.7 eV) being chosen so as to align the Fermi level of Ni to
the Fermi level for gold (-10 eV) obtained using the extended
H\"{u}ckel parameters.  Again the off-diagonal matrix elements of the
tight-binding Hamiltonian for Ni are adjusted accordingly.  We treat
the molecule using extended H\"{u}ckel. As was done previously for the
Ni break-junction system, the coupling parameters between Ni and
molecule are evaluated using extended H\"{u}ckel.

Transmission probabilities for the situation where the substrate is
gold are shown in Fig.~\ref{fig6}. Fig.~\ref{fig6}a is for the case
where the STM tip is made of gold; we include it as a reference for
comparison with our results for the more complicated magnetic case
where the STM tip is made of nickel. The features seen in $T(E)$ in
Fig.~\ref{fig6}a are qualitatively similar to what was seen for the
break-junction system. The resonance at $-8$~eV is due to the
LUMO of BT. There is an overall rise in transmission below $-10$~eV which is
due to the hybridized HOMO states.

The calculated $T(E)$ for the case where the tip is made of Ni are
shown in Fig.~\ref{fig6}b,c: (b) shows $T_{down}(E)$ for the spin down
electrons. (c) shows $T_{up}(E)$ for the spin up electrons.  Again,
the d-states from the Ni hybridize with the molecular states to
provide channels for transport. There are striking differences between
the two curves below $-10$ eV (the adjusted Fermi energy of the Ni
tip). Because the spin down d-band straddles the Fermi energy there are
d-state resonances near $-10$ eV. For the spin up electrons, the
d-band begins about 1~eV below this, and so no spin up d-states
contribute to transport within this window of energy. Because of this
difference, there is a net difference between the number of spin down
versus spin up electrons transmitted into the substrate. This
generates a non-equilibrium population of spin states in the s-band of
the gold substrate. To quantify this difference we have calculated the
current due to spin up and spin down electrons in this system as a
function of applied bias. This is shown in Fig.~\ref{fig7}. The spin
down current is greater than the spin up current by a factor of
approximately 2 to 4. Thus a strong spin polarization of the
current should be expected in this system.

The last case that we consider is that where both the tip and
substrate are made of nickel. Here, we are again interested in whether
spin-valve behavior should be observable. Since the only metal present
in the system is Ni we revert to the tight binding and Fermi energy
parameters that were described in Sec.~4.  The transmission
probability for this system where the magnetizations of the substrate
and tip are aligned parallel and anti-parallel are shown in
Fig.~\ref{fig8}(a) and (b) respectively. For the parallel case there
is a resonance near $-8.7$~eV which is strongly suppressed in when the
magnetizations are anti-parallel. This should manifest itself as a
noticeable difference in differential conductance. Fig.~\ref{fig9}
shows the calculated differential conductance at room temperature,
again using a Fermi energy of $-8.5$~eV for nickel\cite{asym}. The
spin valve effect is now more prominent than that predicted for the
break-junction system where the difference between the strengths of
the resonances in the parallel and antiparallel configurations is not
as large.  The first rise in conductance for the parallel aligned
configuration in Fig.~\ref{fig9} is due to transport through the
resonance at $-8.7$ eV. This resonance is much weaker in the
anti-aligned configuration, as is the corresponding feature in the
conductance in this case. The difference in conductance between the
two cases is now a factor of two, which should be readily detectable.

\section{Conclusions}

In conclusion, we have shown that spin-dependent transport should be
readily observable in molecular wire systems. First, we gave a summary
of a model of electrical conduction through self-assembled molecular
mono-layers that bridge metal break-junctions \cite{Emberly5}. We
assumed that current flows through overlapping molecules each of which
is chemically bonded to only one tip. We showed that this model can
quantitatively and qualitatively reproduce the experimental data
reported by Reed {\em et al.} \cite{Reed1} for a non-magnetic system.
Using this model, we then proceeded to show that it should be possible
to observe spin-valve behavior in a break-junction formed from
ferromagnetic nickel and bridged by SAM's of 1,4 benzene-dithiol. The
spin-valve behavior occurs because of band mismatch between the two
ferromagnetic leads, with the molecules providing transporting states
between the two leads. The second class of systems we considered was a
variety of STM configurations, with the STM tip made from
single-crystal and single-domain nickel.  We showed that it should be
possible to inject a highly spin-polarized current into a gold
substrate which has a self-assembled monolayer of benzene-thiol on
it. Lastly, we showed that spin-valve behavior should be detectable if
the substrate is nickel covered by a SAM of benzene-thiol. We found
that the STM set-up produced a larger spin valve effect than the
break-junction system. These calculations show that semi-empirical
transport methods can be useful in exploring phenomenology of
molecular wires, and yield results that should be helpful in
understanding which effects play an important role in governing
electron transport.

We would like to thank Ross Hill, Bret Heinrich, and Alistair Rowe
for rewarding discussions and helpful suggestions. This work was
supported by NSERC and by the Canadian Institute for Advanced
Research.

\section{Appendix}
Here we review how to change the Hamiltonian matrix elements when the
energy is rescaled and a constant energy shift is added to the
scattering problem for the case where the underlying basis is
non-orthogonal. Schroedinger's equation for the scattering problem is
$H |\Psi\rangle = E|\Psi \rangle$.  We can scale this equation by a
factor $c$, and add and subtract a constant energy shift $\epsilon$
yielding
\begin{equation}
c H |\Psi\rangle = c E |\Psi \rangle + \epsilon |\Psi\rangle -
\epsilon |\Psi\rangle.
\end{equation}
We redefine the scattering energy to be $E' = c E - \epsilon$. Using a
non-orthogonal basis the above equation becomes a matrix equation,
\begin{equation}
\sum_j (c H_{ij} - \epsilon S_{ij}) \Psi_j = E' \sum_j S_{ij} \Psi_j.
\end{equation}
We can define a new Hamiltonian matrix after the application of
a scaling and a constant energy shift by
\begin{equation}
H'_{ij} = c H_{ij} - \epsilon S_{ij}.
\end{equation}
Thus we see that not only are the diagonal elements changed by the
shift, but also the off-diagonal elements when the basis is
non-orthogonal.

%\begin{references}

%\end{references}
%\end{multicols}

%FIGURES
\begin{figure}
%\psfig{file=figs/fig1.eps,width=0.8\textwidth}
%\includegraphics[bb = 0 40 580 400, width =0.85\textwidth,clip]{figs/fig1.eps}
%\includegraphics[bb = 0 0  640 800,clip,width =
%0.75\textwidth]{fig1.eps}
\caption{Schematic of two molecular wire configurations. (a) Metal
break-junction bridged by self-assembled monolayers of organic
molecules. (b) Clean STM tip approaches a self-assembled molecular
monolayer of organic molecules on a metal substrate.}
\label{fig1}
\end{figure}
\begin{figure}
%\psfig{file=figs/fig2.eps, width=0.9\textwidth,angle=-90}
%\includegraphics[bb = 0 0 576 689, width
%=0.85\textwidth,angle=-90,clip]{figs/fig2.eps}
%\includegraphics[bb = 0 0  640 800,clip,width =
%0.75\textwidth]{fig2.eps}
\caption{(a) Calculated average transmission probability at zero applied
bias over 50 different atomic configurations for the system
shown in Fig.~\ref{fig1}a, where the break-junction is made of gold.
(b) Solid line: Calculated differential conductance at room
temperature using a Fermi energy of the gold leads of
$-10$~eV. Dashed line: Experimental differential conductance
from M.~A.~Reed, C.~Zhou, C.~J.~Muller,
T.~P.~Burgin, and J.~M.~Tour, Science {\bf 278}, 252 (1997).}
\label{fig2}
\end{figure}
\begin{figure}
%\psfig{file=figs/fig3.eps,width=0.9\textwidth,angle=-90,clip}
%\includegraphics[bb = 0 0 576 689, width
%=0.85\textwidth,angle=-90,clip]{figs/fig3.eps}
%\includegraphics[bb = 0 0  640 800,clip,width =
%0.75\textwidth]{fig3.eps}
\caption{Simplified schematic of the band structure of ferromagnetic nickel. On
the left, the band structure of spin down electrons. The d-band is
partly empty. On the right, the band structure for the spin up
electrons. Here, the d-band is completely occupied. The 0.54e hole in
the d-band of the spin down electrons gives Nickel a net magnetic
moment of 0.54$\mu_B$.}
\label{fig3}
\end{figure}
\begin{figure}
%\psfig{file=figs/fig4.eps,width=0.9\textwidth,angle=-90,clip}
%\includegraphics[bb =0 0 576 689, width
%=0.85\textwidth,angle=-90,clip]{figs/fig4.eps}
%\includegraphics[bb = 0 0  640 800,clip,width =
%0.75\textwidth]{fig4.eps}
\caption{Transport properties for a single configuration of an
overlapping pair of BDT molecules bonded to a Ni break-junction. (a)
Total transmission probability for the case where the magnetizations on
the left and right leads are aligned parallel. (b) Total transmission
probability for the case where the magnetizations are aligned
anti-parallel. The vertical line indicates the chosen Fermi energy for
the nickel leads of $-8.5$~eV (see text).}
\label{fig4}
\end{figure}
\begin{figure}
%\psfig{file=figs/fig4.eps,width=0.9\textwidth,angle=-90,clip}
%\includegraphics[bb =0 0 576 689, width
%=0.85\textwidth,angle=-90,clip]{figs/fig4.eps}
%\includegraphics[bb = 0 0  640 800,clip,width =
%0.75\textwidth]{fig4.eps}
\caption{Calculated differential conductances of the Ni-BDT
break-junction system at room temperature using a Fermi energy of
$-8.5$~eV. (Solid line) case where the magnetizations of the two tips
are aligned parallel. (Dashed line) case where the magnetizations are
aligned anti-parallel.}
\label{fig5}
\end{figure}
\begin{figure}
%\psfig{file=figs/fig4.eps,width=0.9\textwidth,angle=-90,clip}
%\includegraphics[bb =0 0 576 689, width
%=0.85\textwidth,angle=-90,clip]{figs/fig4.eps}
%\includegraphics[bb = 0 0  640 800,clip,width =
%0.75\textwidth]{fig4.eps}
\caption{Transmission probabilities for the STM set-up using a gold
substrate and benzene-thiol as the molecular wire. (a) case where the
STM is gold. (b) Transmission probability for spin down electrons for
the Ni tip. (c) Transmission probability for spin up electrons for the
Ni tip.}
\label{fig6}
\end{figure}
\begin{figure}
%\psfig{file=figs/fig4.eps,width=0.9\textwidth,angle=-90,clip}
%\includegraphics[bb =0 0 576 689, width
%=0.85\textwidth,angle=-90,clip]{figs/fig4.eps}
%\includegraphics[bb = 0 0  640 800,clip,width =
%0.75\textwidth]{fig4.eps}
\caption{Calculated current at room temperature as a function of
applied bias for the Ni STM tip - BT - Au substrate system. (Solid line)
current due to spin down electrons. (Dashed line) current due to spin
up electrons.}
\label{fig7}
\end{figure}
\begin{figure}
%\psfig{file=figs/fig4.eps,width=0.9\textwidth,angle=-90,clip}
%\includegraphics[bb =0 0 576 689, width
%=0.85\textwidth,angle=-90,clip]{figs/fig4.eps}
%\includegraphics[bb = 0 0  640 800,clip,width =
%0.75\textwidth]{fig4.eps}
\caption{Transmission probabilities for the case where the substrate
is nickel. (a,b) Total transmission probability for the case where the
substrate and tip have their magnetizations aligned parallel and
anti-parallel respectively.}
\label{fig8}
\end{figure}
\begin{figure}
%\psfig{file=figs/fig4.eps,width=0.9\textwidth,angle=-90,clip}
%\includegraphics[bb =0 0 576 689, width
%=0.85\textwidth,angle=-90,clip]{figs/fig4.eps}
\caption{Calculated differential conductance at room temperature for
the Ni STM tip - BT - Ni substrate system, using a Fermi energy of
$-8.5$~eV. (Solid line) STM tip and substrate magnetizations aligned
parallel, (dashed line) aligned anti-parallel.}
\label{fig9}
\end{figure}


\begin{thebibliography}{00}
\bibitem{so}In certain semiconductor systems it is expected that
spin-orbit coupling should give rise to interesting spin transport
phenomena\cite{Datta-Das,Mireles-Kirczenow} through the Rashba
mechanism\cite{Bychkov-Rashba}, however in this work we are
concerned with small molecules
consisting of relatively light atoms that are
not expected to exhibit pronounced spin-orbit-related
transport  effects.
\bibitem{Transmission} We found that the transmission probability
was larger for the energetically unfavorable configuration where the two
molecules are aligned parallel compared to the more favorable perpendicular
alignment.
\bibitem{asym}We have assumed symmetric drops in potential
at the two metal-molecule interfaces
in the Landauer formula, and have not calculated $T(E)$ at higher
applied biases. However, the system is asymmetrically coupled and
therefore some asymmetry in $dI/dV$ may be expected  to be observed
experimentally as the STM tip is retracted from the molecule and the
asymmetry between the couplings at the two molecule-metal interfaces
becomes more pronounced.
\end{thebibliography}
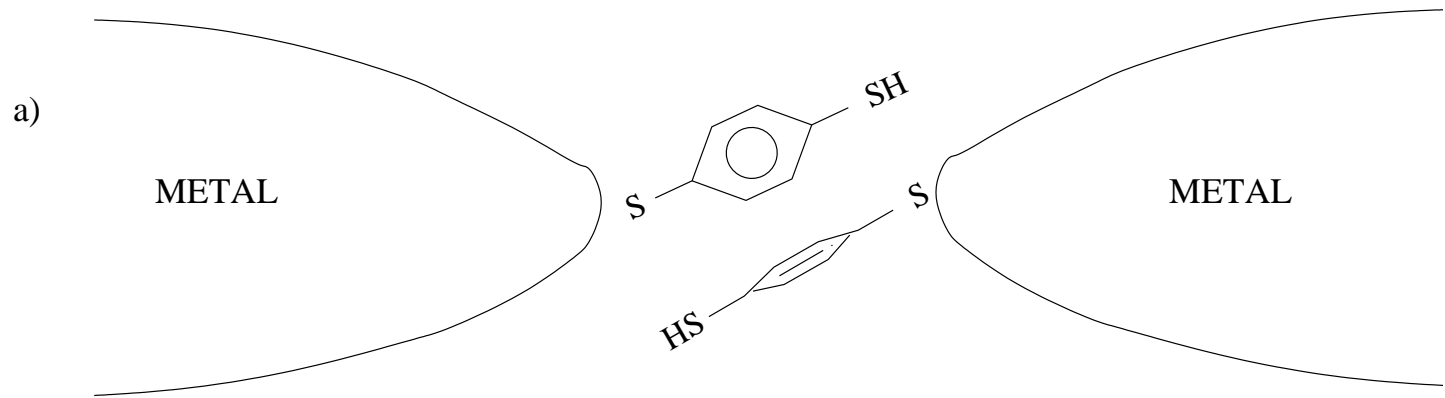
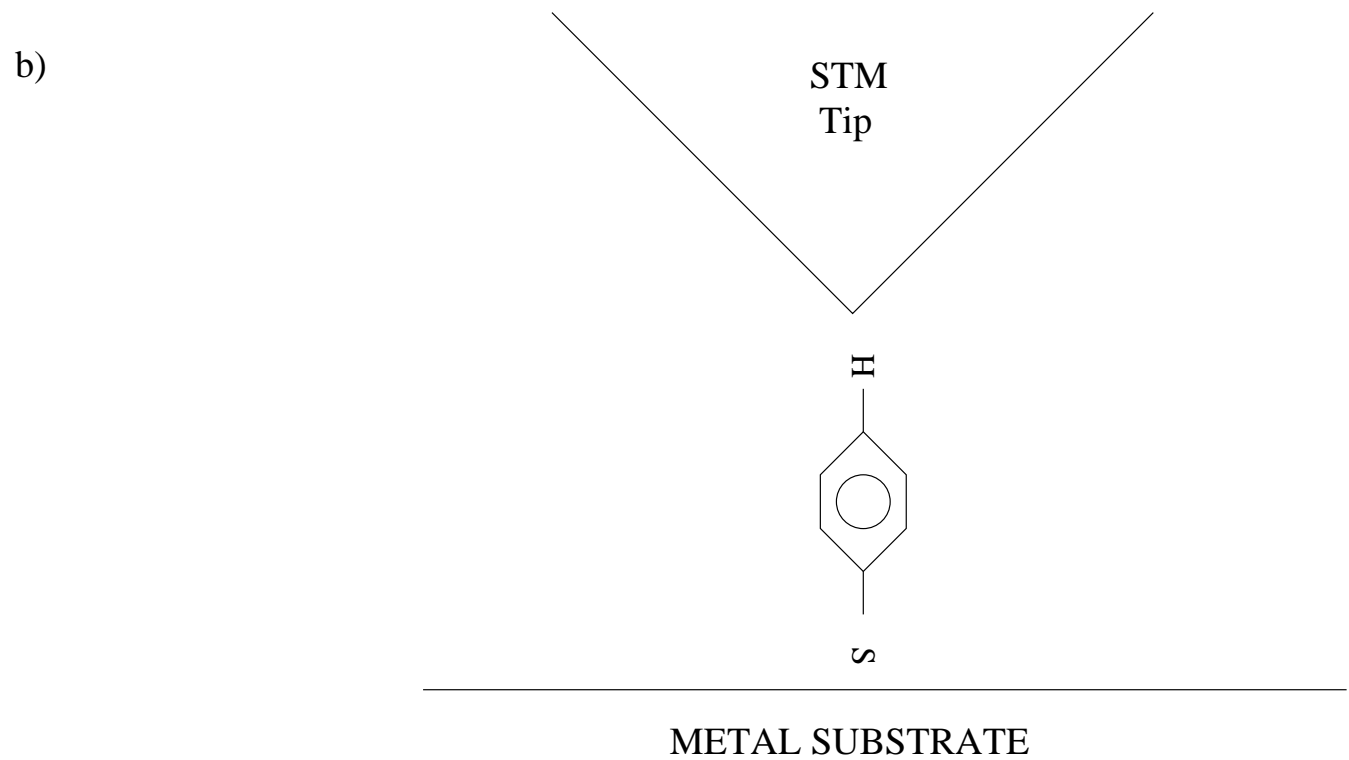
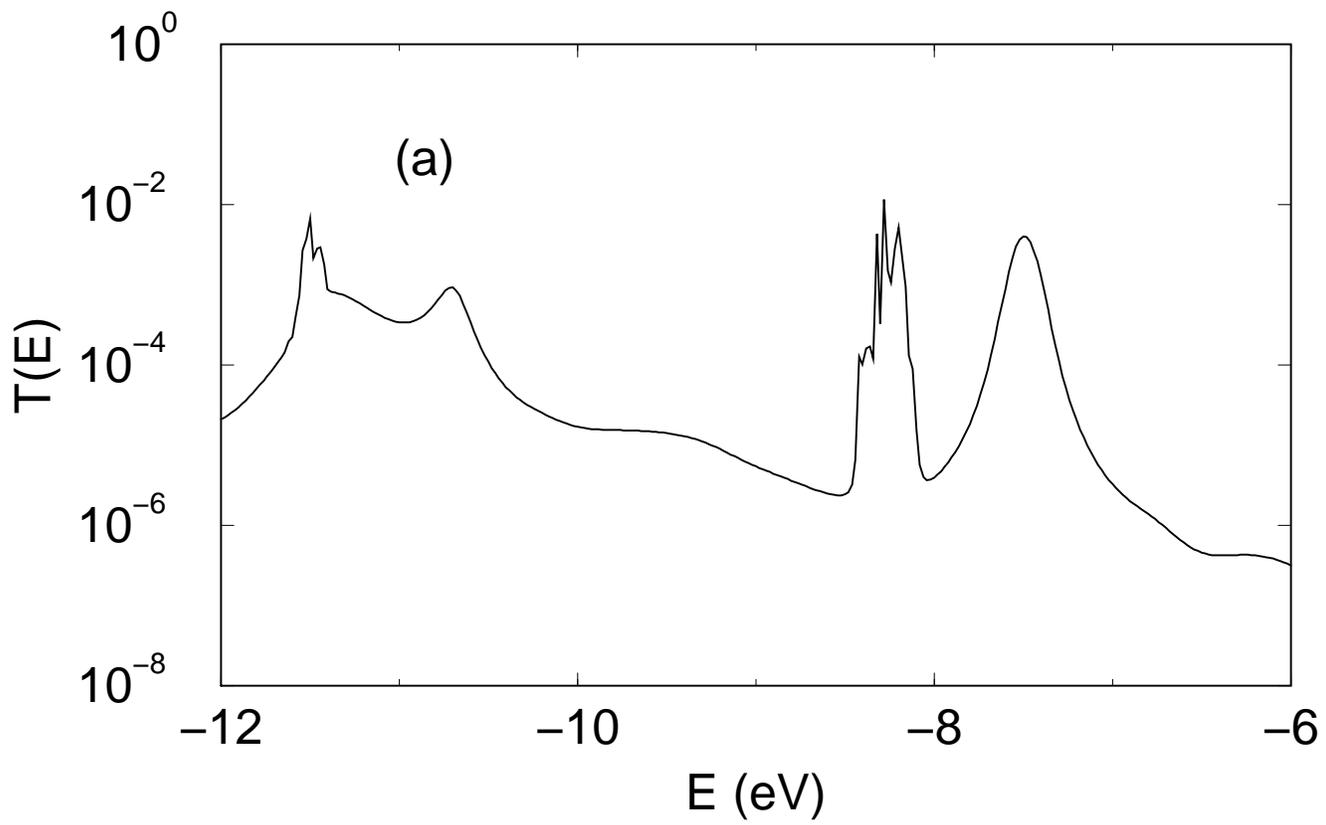
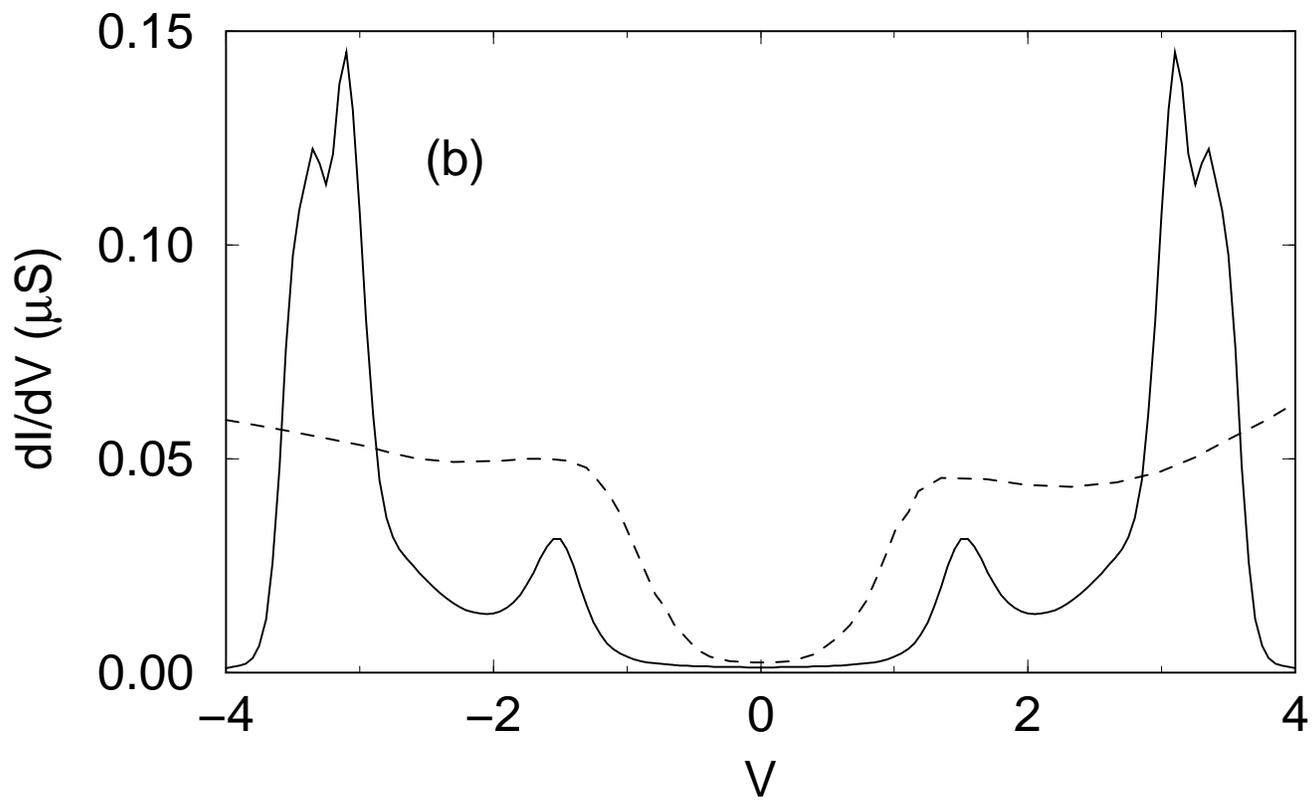
\end{document}